\documentclass[pra,aps,superscriptaddress,11pt]{revtex4-2}
\usepackage[utf8]{inputenc}
\usepackage[english]{babel}
\usepackage[T1]{fontenc}
\usepackage{textgreek}
\DeclareUnicodeCharacter{03BC}{\textmu}
\usepackage{amsmath}
\usepackage{hyperref}
\usepackage{bm} 
\usepackage{amssymb}
\usepackage{graphicx}
\usepackage{xcolor}

\begin{document}
\title{Enhanced quantum sensing of gravitational acceleration constant}
\author{Giorgio Stucchi}
\email{giorgio.stucchi@tum.de}
\affiliation{Max Planck Institute of Quantum Optics, Hans-Kopfermann-Str. 1, Garching 85748, Germany}
\author{Matteo G. A. Paris}
\email{matteo.paris@fisica.unimi.it}
\affiliation{Dipartimento di Fisica, Universit\`a di Milano, I-20133 Milan, Italy}
\begin{abstract}
We investigate the use of quantum probes to accurately determine the strength of the local gravitational field on Earth. Our findings show that delocalized probes generally 
outperform localized ones, with the precision enhancement scaling quadratically with 
the separation between the two wavefunction components.
This advantage persists under realistic position measurements, which can achieve 
precision not too far from the ultimate bound. 
We also discuss the influence of Earth's surface, demonstrating that its effect 
can be neglected until shortly before the particle hits the floor. Finally, we 
address the joint estimation of the gravitational acceleration $g$ and the probe 
mass $m$, proving that the excess estimation noise arising from their inherent 
incompatibility is negligible.
\end{abstract}
\maketitle
\section{Introduction}\label{Introduction}
Over the past fifty years, there has been significant interest in exploring the relationship between quantum mechanics and gravity \cite{amelino2005three,sep-quantum-gravity,ciufolini2016test,PhysRevD.99.104029,PhysRevD.110.024022,PhysRevD.109.064073}. In the non-relativistic weak field regime, precise tests have been conducted to validate predictions based quantum mechanics and a Newtonian classical field. These experiments have successfully demonstrated that the dynamics of a quantum probe in a uniform gravitational field can be effectively described by a Schrödinger equation with a Hamiltonian incorporating a linear gravitational term as potential energy. This approach has enabled the prediction and measurement of gravitational effects on quantum systems using relatively simple experimental setups \cite{Carlesso2020,Biswas2023,Chen2024}. 
One of the pioneering experiments in this field was the COW experiment \cite{COW}, which measured a gravitational effect in a quantum system. The experiment observed a phase-shift induced by gravity in a Mach-Zehnder interferometer using thermal neutrons. If the two arms of the interferometer are positioned at different heights, the wave-function accumulates a gravity-induced phase-shift that is absent when the interferometer is placed horizontally.

More recently, collaborations such as qBounce \cite{ABELE2014,jenke2019testing} and GRANIT 
\cite{Nesvizhevsky2012} have investigated ultracold neutrons confined by a 
perfect mirror. These neutrons serve as ideal probes due to their light mass 
and null electric charge. In addition to the gravitational potential, the particles 
also experience an infinite barrier at the origin. The dynamics of these quantum 
bouncers have been explored theoretically and experimentally. Ultracold neutrons 
provide a very useful tool to investigate gravitational phenomena in the quantum regime \cite{PhysRevD.93.122006,PhysRevD.109.064085}, and 
can be used as test of interactions beyond the Standard Model and General Relativity 
in the qBOUNCE and GRANIT experiments, for example as a tool for probing of 
beyond-Riemann gravity \cite{Ivanov2021}, or influence of a spontaneous Lorentz 
symmetry breaking on the gravitational quantum states of the ultracold neutrons 
\cite{Escobar2022}. Experiments have also been proposed and conducted to examine 
the free fall of antimatter particles \cite{sala2015matter}, as the GBAR experiment \cite{perez2015gbar}.

The coupling of a quantum probe to gravity also gives the opportunity to estimate 
the gravitational acceleration, which in turn may provide constraints on any possible 
gravity-like interaction at a given interaction range \cite{SISSA2015,PhysRevD.107.064055}. Indeed, 
the phase difference accumulated in an interferometer \cite{peters2001high}, or 
the discrete energy spectrum of a quantum bouncer \cite{gea1999quantum} do depend 
on the value of $g$. We thus aim to determine the fundamental limits to the precision 
in the estimation of $g$ \cite{PhysRevResearch.7.013016,PhysRevA.99.032350}, 
and how it is possible to achieve such ultimate bounds. 
In particular, we want to assess the use of superpositions of states rather than 
localized ones. 

Our findings show that delocalized probes generally 
outperform localized ones, with the precision enhancement scaling quadratically with 
the separation between the two wavefunction components.
This advantage persists under realistic position measurements, which can achieve 
precision not too far from the ultimate bound. Concerning the influence of Earth's 
surface, we prove that its effect 
can be neglected in a first approximation. We also address the joint estimation 
of the gravitational acceleration $g$ and the mass $m$ of the quantum probe, proving 
that the no excess estimation noise is arising, despite their inherent incompatibility 
as parameters. 

The paper is structured as follows. In Sections \ref{Estimation Theory} and \ref{evol} 
we briefly review the tools of quantum estimation theory and the quantum 
description of a free falling body and a quantum bouncer, respectively. In Section \ref{qetg}
we derive the ultimate quantum bound to precision in the estimation of $g$ for a localized probe as well as for superpositions, showing that delocalized probes provide a 
precision enhancement scaling quadratically with the separation between the two 
wavefunction components. In Section \ref{bounc} we prove that there is a large temporal 
range in which the influence of Earth's surface
can be neglected, validating the results of Section \ref{qetg}. 
In Section \ref{position} we study the performance of position measurements, 
showing that the advantage of superpositions persists, and that the Fisher information 
is not {much} smaller than the QFI. Finally, in Section \ref{joint} we discuss 
joint estimation of the gravitational acceleration and the probe mass, proving 
that the excess estimation noise arising from their inherent incompatibility 
is negligible. Section \ref{outro} closes the paper with some concluding remarks.
\section{Classical and Quantum Estimation Theory}\label{Estimation Theory}
In classical estimation theory, the central challenge lies in estimating an unknown parameter, denoted as $\lambda$, given a set of outcomes $x_1, x_2, \ldots, x_M$ from a set of experimental data distributed according to a conditional probability distribution $p(x| \lambda)$. Given $\lambda$ and a sample space $\chi=\{x\}$, namely a subset of all possible outcomes of the experiment, an estimator $\hat{\lambda}$ is a function from the sample space $\chi$ to the space of the possible values of $\lambda$. 
The variance $\mathrm{Var}(\hat{\lambda})$ of any unbiased estimator $\hat{\lambda} =\hat{\lambda}(x_1, x_2, ..., x_M )$ satisfies the Cramér-Rao inequality \cite{MMS}:
\begin{equation}\label{eqn:Cramer-Rao} 
 \mathrm{Var}(\hat{\lambda}) \ge \frac{1}{M F(\lambda)},
\end{equation}
where $F(\lambda)$ is the Fisher Information (FI) defined as: 
\begin{equation}\label{eqn:Fisher} 
F(\lambda)=\int\! dx \, p(x|\lambda) \left(\frac{\partial \log(p(x|\lambda))}{\partial\lambda}\right)^2.
\end{equation}
According to the Cramér-Rao bound the optimal measurement to estimate the quantity $\lambda$ is the one with conditional distribution $p(x|\lambda)$ that maximizes the Fisher Information. On the other hand, for any fixed measurement an efficient estimator is an estimator that saturates the Cramér-Rao inequality \cite{QEQT}.

For quantum systems, any estimation problem can be formulated by considering a family of quantum states $\rho_{\lambda}$ defined on a specific Hilbert space $\mathcal{H}$. To estimate the value of $\lambda$, measurements are performed on the quantum state $\rho_{\lambda}$ followed by classical data processing on the measurement results 
\cite{QDE,PSQT}. 
Introducing the Symmetric Logarithmic Derivative (SLD) $L_{\lambda}$ as the
self-adjoint operator satisfying the equation
$\frac12(L_{\lambda}\rho_{\lambda}+\rho_{\lambda}L_{\lambda}) = \partial_\lambda \rho_{\lambda}$, the following inequality can be obtained:
\begin{equation}\label{eqn:QuantumCramerRao} 
 \mathrm{Var}(\hat{\lambda}) \ge \frac{1}{M H(\lambda)},
\end{equation}
where the Quantum Fisher Information (QFI) is defined as $H(\lambda) = \mathrm{Tr}[\rho_{\lambda} L_{\lambda}^2] $ and is always greater than or equal to the FI of any measurement procedure. The spectral measure of the SLD represents the optimal measurement to be performed on the system, i.e. the measurement with a FI equal to the QFI.

For a generic family of pure states the QFI is given by
\begin{equation}\label{eqn:qfipurestate}
H(\lambda) = 4 [\langle \partial_\lambda \psi_{\lambda} |  \partial_\lambda\psi_\lambda \rangle  -\left|\langle \partial_\lambda\psi_{\lambda}  | \psi_{\lambda} \rangle\right|^2].
\end{equation}
If the parameter is encoded on the quantum system via a parameter-dependent unitary transformation $U$, we introduce the Hermitian operator
\begin{equation}\label{eqn:curlyGdefinition}
\mathcal{G} := i (\partial_\lambda U^\dagger) U\,,
\end{equation}
and the QFI can be expressed as \cite{QMUP}
\begin{equation}\label{eqn:qfiunitary}
H(\lambda) = 4\left[ \langle\psi_0|\mathcal{G}^2|\psi_0\rangle -\langle\psi_0|\mathcal{G}|\psi_0\rangle^2 \right]\,.
\end{equation}
In other words, the QFI is proportional to the
variance of $\mathcal{G}$ on the initial state $|\psi_0 \rangle$. For a parametrization transformation of the form $U = \mathrm{exp}(-itJ_\lambda)$, where $\hbar$ has been set as 1, $\mathcal{G}$ can be expressed as
 \begin{equation}\label{eqn:curlyG}
\mathcal{G} = -\int_0^t\! ds \, e^{i s J_\lambda} (\partial_\lambda J_\lambda) e^{- i s J_\lambda}.
\end{equation}

\subsection{Multiparameter Quantum Estimation}\label{Multiparameter Quantum Estimation}
The dynamics of a system are often influenced by multiple interconnected parameters. However, extending results such as the Cramér-Rao bound to the multiparameter case is not straightforward, because the optimal measurement strategy for one parameter may not commute with the optimal scheme for another parameter. This incompatibility between measurements arises from fundamental principles in quantum mechanics. Furthermore, when a measurement is fixed, correlations can emerge among the parameters. Consequently, trade-off relations among the uncertainties of the parameters arise, as the initial information in the measurement probe needs to be distributed among the different parameters \cite{PMQM,liu2020quantum}.

Similarly to the single parameter case, symmetric logarithmic derivative (SLD) 
operators $L_\mu$ may be introduced for each parameter, and the so-called QFI 
matrix  may be constructed, with elements
\begin{equation}
 H_{\mu\nu}(\bm{\lambda}) =  \mathrm{Tr}\left[\rho_{\bm{\lambda}} \frac{L_\mu L_\nu + L_\nu L_\mu}{2}\right]\,,
\end{equation}
leading to a matrix quantum Cramér-Rao bound $
\bm{V(\hat{\lambda})} \ge \bm{H(\bm{\lambda})}^{-1}$
where $\bm{V}$ is the mean square error matrix (the covariance matrix 
for unbiased estimators). This relation is of little use, since no ordering
is avalaible for matrices.
To get a better insight into the performance of different multiparameter estimators, 
it is convenient to introduce a weight matrix $\bm{W}$
(positive, real matrix of dimension $d \times d$) and recast the
matrix bound into a scalar bound as follows
\begin{equation} \label{SLD-CRB scalar}
 \mathrm{Tr}\left[ \bm{ W V} \right] \ge \mathrm{Tr} \left[ \bm{W}\bm{H}^{-1} \right] 
 \equiv C^S(\bm{W})\,.
\end{equation}
This bound is in general not attainable, due to the 
possible non-commutativity of the different SLD operators \cite{PMQM}.  
Holevo derived a tighter scalar bound $C^H(\bm{\lambda}, \bm{W})$ 
via explicit minimization 
\cite{PSQT}, which can be achieved by collective measurements on 
an asymptotically large number of copies of the state.
Closed form expressions of the Holevo bound are hard to obtain, but
it can be upper- and lower bounded as follows:
\begin{equation}
\label{CH-bounds}
 C^S(\bm{W}) \le C^H \le (1+\mathcal{R})\,C^S(\bm{W}) \,,
\end{equation}
where the quantity $\mathcal{R}$ is defined as
\begin{equation}\label{eqn:Rdef}
\mathcal{R} = \lVert i \bm{H}^{-1} \bm{D} \rVert_\infty = \sqrt{\frac{\det \bm{D}}{\det \bm{H}}}
\end{equation}
where $\lVert \bm{A} \rVert_\infty$ denotes the largest eigenvalue of the matrix $\bm{A}$, 
and the second equality is valid only for statistical models with two parameters. 
The  matrix $\bm{D}$ is the Uhlmann curvature, with elements
\begin{equation}
D_{\mu\nu} = - \frac{i}{2} \mathrm{Tr} \left[ \rho_\lambda 
\left[ L_\mu^S,  L_\nu^S \right]\right]\,.
\end{equation}
The quantity $\mathcal{R}$ satisfies the constraint $ 0 \leq \mathcal{R} \leq 1$, and 
provides a general bound, independent on the choice of the weight matrix. If a specific 
weighting is more appropriate for the statistical model at hand, a more refined 
bound may be obtained as follows 
\begin{equation}\label{T-bound}
     C^S(\bm{W}) \le C^H \le \left[1+\mathcal{T}(\bm{W})\right]\,C^S(\bm{W}) \,,
     \end{equation}
where 
\begin{equation}\label{eqn:Tdef}
\mathcal{T}(\bm{W}) = \frac{
\lVert \sqrt{\bm{W}} \bm{H}^{-1} \bm{D} \bm{H}^{-1} \sqrt{\bm{W}} \rVert_1}{C^S(\bm{W})} 
= \frac{2 \sqrt{w \det \bm{D}}}{H_{11} + w H_{22}}
\end{equation}
with $\lVert \bm{A} \rVert_1$ denoting the sum of the singular values of the matrix $\bm{A}$, 
and the second equality valid only for statistical models with two parameters and diagonal 
weight matrix $\bm{W} = \hbox{Diag} (1, w)$.
\section{Falling quantum particles}\label{evol}
\subsection{Quantum free fall}
The idealized system is given by a quantum particle evolving in a uniform gravitational field, i.e.\ in a potential $V(x) = mgx$. This coincides with the first order expansion of the Newtonian potential $\tilde{V}(x)=-\frac{G M m}{(R+x)^2} $ after substitution of $g = \frac{G M}{R^2} $,  having denoted with $x$ the distance from the floor, with $M$ and $R$ the mass and radius of the Earth respectively, and with $G$ the gravitational constant. The approximation consists in neglecting the presence of the floor. The Hamiltonian of the system is given by \cite{wadati1999free,PhysRevLett.131.010801}
\begin{equation}
    \mathcal{H} = \frac{p^2}{2m} + mg{x}.
\end{equation}
The t-independent Schrödinger equation reads:
\begin{equation}\label{Schrodinger}
   \left\{ - \frac{\hbar}{2m} \left( \frac{d}{dx}\right)^2 + mgx\right\} \Psi(x) = E\Psi(x)
\end{equation}
Introducing the shifted and re-scaled new independent variable $y = k(x - \frac{E}{mg})$ and adopting the notation $\Psi(x)=\sqrt{k} \psi(y)$, with $k = (\frac{2m^2g}{\hbar^2})^\frac{1}{3}$, the differential equation then simply becomes 
\begin{equation}\label{eqn:Airy}
   \frac{d^2}{dy^2}\psi(y) = y\psi(y)
\end{equation}
Eq.~(\ref{eqn:Airy}) is Airy’s differential equation: its solutions are linear combinations of the $\textit{Airy functions } \mathrm{Ai}(y)$ and $\mathrm{Bi}(y)$, of which (since
$\mathrm{Bi}(y)$ diverges as $y\rightarrow\infty$) only the former
\begin{equation}
   \mathrm{Ai}(y) \equiv \frac{1}{\pi} \int_0^\infty du \cos{\left(yu + \frac{1}{3}u^3\right)} 
\end{equation}
are selected.
These solutions are only improperly normalizable: by imposing that
\begin{equation}
    \int_{-\infty}^\infty dx \, \psi(y) \psi^*(y) = \delta(E - E')
\end{equation}
one finds the final expression for the energy eigenfunctions \cite{AIRY}:
\begin{equation}
   \Psi_E(x) = \frac{(2m)^\frac{1}{3}}{\hbar^\frac{2}{3}(mg)^\frac{1}{6}} \mathrm{Ai}\left(k\left(x-\frac{E}{mg}\right)\right).
\end{equation}
The energy spectrum is continuous, and has no least
member: the system possesses no ground state. The eigenfunctions $\Psi_E(x)$ comprise a complete orthonormal set \cite{wheeler}. With them, the propagator can be constructed
\begin{equation}\label{propagator}
    K(x_1, t_1; x_0, t_0) = \int_{-\infty}^\infty dE \, \Psi_E(x_1) \Psi_E^*(x_0) e^{-\frac{i}{\hbar} E (t_1 - t_0)} 
\end{equation}
which allows one to describe the dynamical evolution of any prescribed initial state:
\begin{align}\label{eqn:map}
|\psi_t \rangle  & = e^{- \frac{i}{\hbar} {\cal H} t} |\psi_0\rangle  \notag \\
|\psi_0\rangle & = \int\!\! dx\, \psi(x, 0)\, |x\rangle \notag \\
|\psi_t\rangle & = \int\!\! dx\, \psi(x, t)\, |x\rangle \notag \\
\psi(x, 0) \mapsto \psi(x, t) & = \int dx_0 \, K(x, t; x_0) \psi(x, 0) 
\end{align}
\begin{widetext}
Eq.~(\ref{propagator}) yields \cite{wheeler}
\begin{equation}\label{propagatorofsysteminexam}
    K(x, t; x_0) = \sqrt{\frac{m}{2 \pi i \hbar t}} \exp\left[ \frac{i}{\hbar}\left(\frac{m}{2 t}(x - x_0)^2 - \frac{1}{2} m g t (x + x_0) - \frac{1}{24} m g^2 t^3\right)\right]
\end{equation}
\end{widetext}
\subsection{Quantum Bouncer}
Let us now consider the presence of the Earth's surface. The potential is obtained by erecting an impenetrable barrier at the coordinate origin, the intended effect
of which is to render inaccessible the points with $x < 0$ \cite{gea1999quantum,nesvizhevsky2000search,doncheski2001expectation,nesvizhevsky2005study}.
\begin{equation}
    V(x) =
\begin{cases}
    \infty & \text{if } x < 0, \\
    mgx & \text{if } x \geq 0.
\end{cases}
\end{equation}
The t-independent Schrödinger equation is the same as for the free fall, but in this case it is required that 
\begin{equation}
\Psi(x<0,t) = 0 \quad \forall t.
\end{equation}
This amounts to a requirement that the probability current $i\frac{\hbar^2}{2m}(\Psi^*\Psi -\Psi\Psi^*)\mid_{x=0} = 0 \quad \forall t$, which is achieved by imposing the boundary condition 
\begin{equation}\label{boundarycondition}
\Psi(0,t) = 0 \quad \forall t.
\end{equation}
The presence of this boundary condition is what sets the two problems apart, and it is a crucial factor. It leads to a discrete energy spectrum, ensures that the eigenfunctions are normalizable, and provides the system with a ground state, which was absent in the case of free fall \cite{wheeler}. Just as in the free fall, the eigenfunctions have the form 
 \begin{equation}
   \Psi(x) \propto \mathrm{Ai}\left(k\left(x- a \right)\right)
\end{equation}
and to achieve compliance with the boundary condition~(\ref{boundarycondition}), $a \equiv \frac{E}{mg}$ must be assigned such a value as to make $-ka$ coincide with a zero, which is denoted as $-z_n$, of the Airy function. Hence $a_n = z_n / k$, and thus the energy eigenvalues are:
\begin{equation}\label{energy eigenvalues}
E_n = \frac{mgz_n}{k} \quad n=1, 2, 3, ...
\end{equation}
The associated eigenfunctions are  
\begin{equation}\label{eqn:eigenfunctions}
   \Psi_n(x) = \mathcal{N}_n \mathrm{Ai}\left(kx- z_n \right)
\end{equation} 
where $\mathcal{N}_n$ is a normalization factor, defined by the condition
\begin{equation}
    \int_0^\infty \left[ \Psi_n(x) \right]^2 dx =  {\mathcal{N}_n}^2 \cdot \int_0^\infty \left[ \mathrm{Ai}\left(kx- z_n \right) \right]^2 dx
\end{equation} 
Hence 
\begin{equation}
    \mathcal{N}_n =  \sqrt{k} \left[ \int_0^\infty \left[ \mathrm{Ai}\left(z- z_n \right) \right]^2 dz \right]^{-\frac{1}{2}}
\end{equation} 
where $z = kx$. This can be simplified further \cite{AIRY}:
\begin{equation}
    \mathcal{N}_n =  \sqrt{k} \left[ \mathrm{Ai'}\left(-z_n \right) \right]^{-1}
\end{equation} 
In this setting, calculating the propagator analytically is unfeasible. Therefore, the dynamically evolved wave packet is conveniently described by
\begin{equation}\label{eqn:seriesexpansion}
    \Psi(x, t) =  \sum_n c_n \Psi_n(x) e^{-\frac{i}{\hbar}E_n t} 
\end{equation} 
\begin{equation}\label{eqn:coefficients}
    c_n =  \int_0^\infty \Psi_n(\xi) \Psi(\xi, 0)  d\xi.
\end{equation} 
\subsection{Static estimation of $g$ using stationary quantum probes}
\label{static}
In recent experiments, eigenstates of the quantum bouncer have been generated {\cite{Nesvizhevsky_2003,Abele_2003}} and a 
question arises on whether they provide a mean to encode and retrieve information on 
$g$. We anticipate that being based on stationary states this estimation strategy 
cannot be improved by dynamical encoding and the QFI is constant in time. Using Eqs.
(\ref{eqn:eigenfunctions}) and (\ref{eqn:qfipurestate}) we arrive at 
\begin{equation}
    H^{(n)}(g) = \frac{1}{{7g^2}} \left(1 + \frac{{32 z_n^3}}{{135}}\right),
\end{equation}
where $-z_n$ is the n-th zero of the Airy function. Preparing eigenstates with 
larger energy would be thus advantageous for the estimation of $g$.
Remarkably, if we compute the FI 
for position measurements, i.e., the FI in Eq.~\eqref{eqn:Fisher} for  the distribution 
$| \psi_n \rangle$ we obtain the same results. 
This means that position measurement is optimal to estimate $g$ when this parameter 
is encoded onto energy eigenstates. In order to obtain a dynamical advantage, i.e., to 
exploit the encoding of $g$ due to evolution, we may consider superpositions of energy 
eigenstates. However, these kind of probes are challenging to produce 
\cite{cronenberg2018acoustic}
and we prefer to
address dynamical estimation for more realistic preparations.
\section{Estimation of $g$ by free falling quantum probes}\label{qetg}
We now analyze whether the parameter $g$ may be estimated by performing measurements 
on a free falling quantum probe, and focus on whether using a superposition of Gaussian 
packets, instead of a localized one, is beneficial from the point of view of extracting 
information from the system. We will be using natural units, with $\hbar = 1$ and $c=1$. 

To calculate the QFI for the parameter $g$ in the no-floor regime, we use formula~(\ref{eqn:qfiunitary}), since we are working with pure states under 
unitary transformations. The initial ($t=0$) state is described by the wavefunction
\begin{align}\label{eqn:singlepacket}
 \psi_0(x) & \equiv \psi(x, 0) = \frac{1}{(2 \pi \sigma^2)^{1/4}} \exp \left(-\frac{x^2}{4\sigma^2}\right) 
    \end{align}
In order to evaluate Eq.~(\ref{eqn:curlyG}) for our system, we make use of the Baker-Campbell-Hausdorff formula and from Eq. (\ref{eqn:curlyG}) we find the operator 
$$\mathcal{G}= -(mt x + \frac{t^2}{2}{p})\,.$$
The QFI is then given by 4 times the variance of $\mathcal{G}$ on the initial state. 
Alternatively, upon observing that the propagator in Eq. (\ref{propagator}) is 
symmetric in its arguments $K(x,t;x_0)=K(x_0,t;x)$ and its derivative
is given by
$$
\partial_g K(x,t;x_0) = - \frac{i}{2} m t \left( \frac{1}{6} g t^2 + x + x_0\right)\, K(x,t;x_0)\,,
$$
the QFI may be expressed in terms of the fluctuations of the position operator 
on the evolved state as
\begin{align} 
 H (g) = 4 m^2 t^2 \Big( \langle\psi_t | x^2 | \psi_t\rangle 
 - \langle\psi_t | x | \psi_t\rangle^2\,.
 \Big) \label{Hdelta}
\end{align}
Either way, for a localized  initial state $|\psi_0 \rangle$ we find
\begin{equation}\label{calc:qfisinglepacket}
    H_{loc} =\frac{t^4}{4\, \sigma^2} + 4\, m^2 t^2 \sigma^2\,,
\end{equation}
which is independent on $g$ itself.
The same results is obtained if the probe has an initial momentum, 
i.e. if the initial wave function is  $\psi_0 (x,0) e^{i p_0 x}$. The QFI increases for decreasing $\sigma$, i.e. if the particle is more localized. The QFI scales as $t^2$ for short times whereas asymptotically (for large times) we have $H_{loc} \propto t^4$. The emergence of this scaling may 
be strongly delayed if $\sigma$ is large, as the first term in Eq.  (\ref{calc:qfisinglepacket}) becomes larger than the second only for $t> 4 m \sigma^2$.

Let us now consider a particle initially prepared in a superposition of 
two Gaussian wavepackets, placed at distance $2 a$ each other, and having 
opposite momenta $\pm p_0$ with $p_0>0$ (i.e. the wavepacket on the right 
has positive momentum and viceversa)
\begin{align}\label{eqn:superposition}
    \psi(x, 0) & = 
    \frac{{e^{-\frac{{(x-a)^2}}{{4 \sigma^2}}+i p_0 x} + e^{-\frac{{(x+a)^2}}{{4 \sigma^2}}-i p_0 x }}}{{A^{\frac12}\,(2\pi \sigma^2)^{\frac{1}{4}} }}   \\
      A & = 2\left(1 + e^{-\frac{{(a^2 + 4 p_0^2 \sigma^4)}}{{2 \sigma^2}}}\right)
\end{align}
In this case, the QFI of the evolved state may be written as
\begin{align}
\label{calc:qfisuperposition}
 H_{sup} = &  H_{loc} + 8\, m^2\, \sigma^2\, t^2 
 \bigg[
\frac{a^2}{2\sigma^2} -  g(a,\sigma, p_0) \bigg]  \\ & + 4\, a\, m\, p_0\, t^3 + \frac{t^4}{2\sigma^2} 
\bigg[2\sigma^2p_0^2 - g(a,\sigma, p_0)\bigg]\,,\notag
 \\
 g(a,\sigma,p_0) & = \frac{
 \frac{a^2}{2\sigma^2} + 2 \sigma^2 p_0^2}
 {1+\exp (\frac{a^2}{2\sigma^2} + 2\sigma^2 p_0^2)}\,.\notag
\end{align}

From Eq. (\ref{calc:qfisuperposition}) it is clear that for $a^2 > 2\sigma^2 \mu$ and 
$p_0^2 > \mu/2\sigma^2$, where $\mu=\max_x e^x/(1+e^x)\simeq 0.278$, i.e. for particles 
that are enough {\em fast and delocalized} compared to the wavepackets width, 
one has $ H_{sup} > H_{loc} $, in agreement with the intuition suggested by Eq. 
(\ref{Hdelta}). For the special cases $a=0$ (initially localized particle, with two 
opposite momenta) and $p_0=0$ (initially delocalized particle with no momentum) we have
\begin{align}
 H_{sup}  =   H_{loc} + & \frac12 t^4 p_0^2\, (1+ \tanh \sigma^2 p_0^2) \notag \\ 
 - & \frac12 t^2 p_0^2\, m^2 \sigma^4\, (1- \tanh \sigma^2 p_0^2) 
 \\
 H_{sup}  =   H_{loc} + & 2 a^2 m^2 t^2 \left(1+ \tanh \frac{a^2}{4\sigma^2}\right) \notag \\ 
+ & \frac{a^2 t^4}{8 \sigma^2}\, \left(1- \tanh\frac{a^2}{4\sigma^2}\right) \label{a2}
\end{align}
The first expression shows that for an initially localized particle with two 
opposite momenta, we gain precision for large times, whereas the second ensures 
that even without an initial momentum, the use of a delocalized probe always 
provides enhanced precision compared to a localized one. The precision 
enhancement scales as the square 
of the separation between the two wavepackets. Notice that for large times, i.e. when the term proportional to $t^4$ in Eq.(\ref{a2}) is dominant, the QFI enhancement is maximized 
for a specific ratio between $a^2$ and $\sigma^2$ i.e. for $a^2 = 4 \xi \sigma^2$ where 
$\xi = \arg \max x (1-\tanh x) \simeq 0.64 $ is the value of $x$ maximizing the function $x (1-\tanh x)$.

Summarizing, the use of quantum probes (superpositions) generally enhances precision 
in the estimation of the strength of the local gravitational field of Earth. Precision 
may be optimized by mild state engineering.

\section{Free falling vs quantum bouncer}\label{bounc}

We now consider the presence of the Earth surface and analyze the dynamics of a {\em 
quantum bouncer} \cite{bc1,bc2,bc3,bc4}. Our goal is to assess whether the {\em no floor} approximation is 
valid for a certain range of time before the particle hits the Earth. We anticipate 
that this is indeed the case, thus justifying the use of the free-falling model, and
validating the results about the QFI obtained in the previous Section. 

The initial states are those given in  Eqs.~(\ref{eqn:singlepacket}) and 
(\ref{eqn:superposition}), the only difference being that the wave packets are now 
translated to take into account the initial height $h$ from the floor. Notice that 
strictly spealing those states are not normalized on $(0, \infty)$. However,  for 
$h \gg \sigma$ (i.e., assuming that the wave packet is dropped from a height large 
compared to its width) they are nearly normalized (and in particular the wavefunction 
vanishes at the Earth surface). 

\begin{figure}[h!]
\centering
\includegraphics[width=0.49\columnwidth]{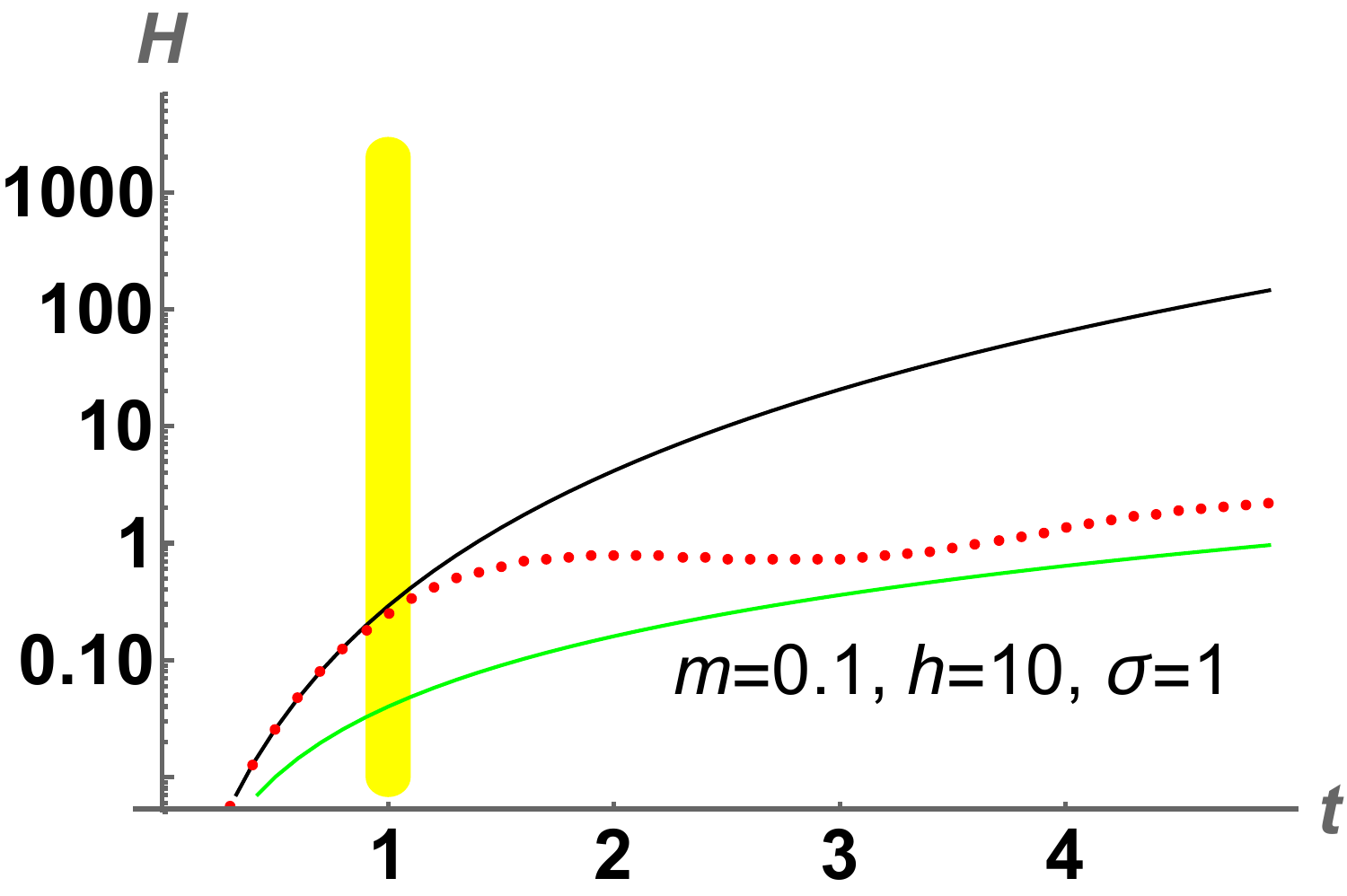}
\includegraphics[width=0.49\columnwidth]{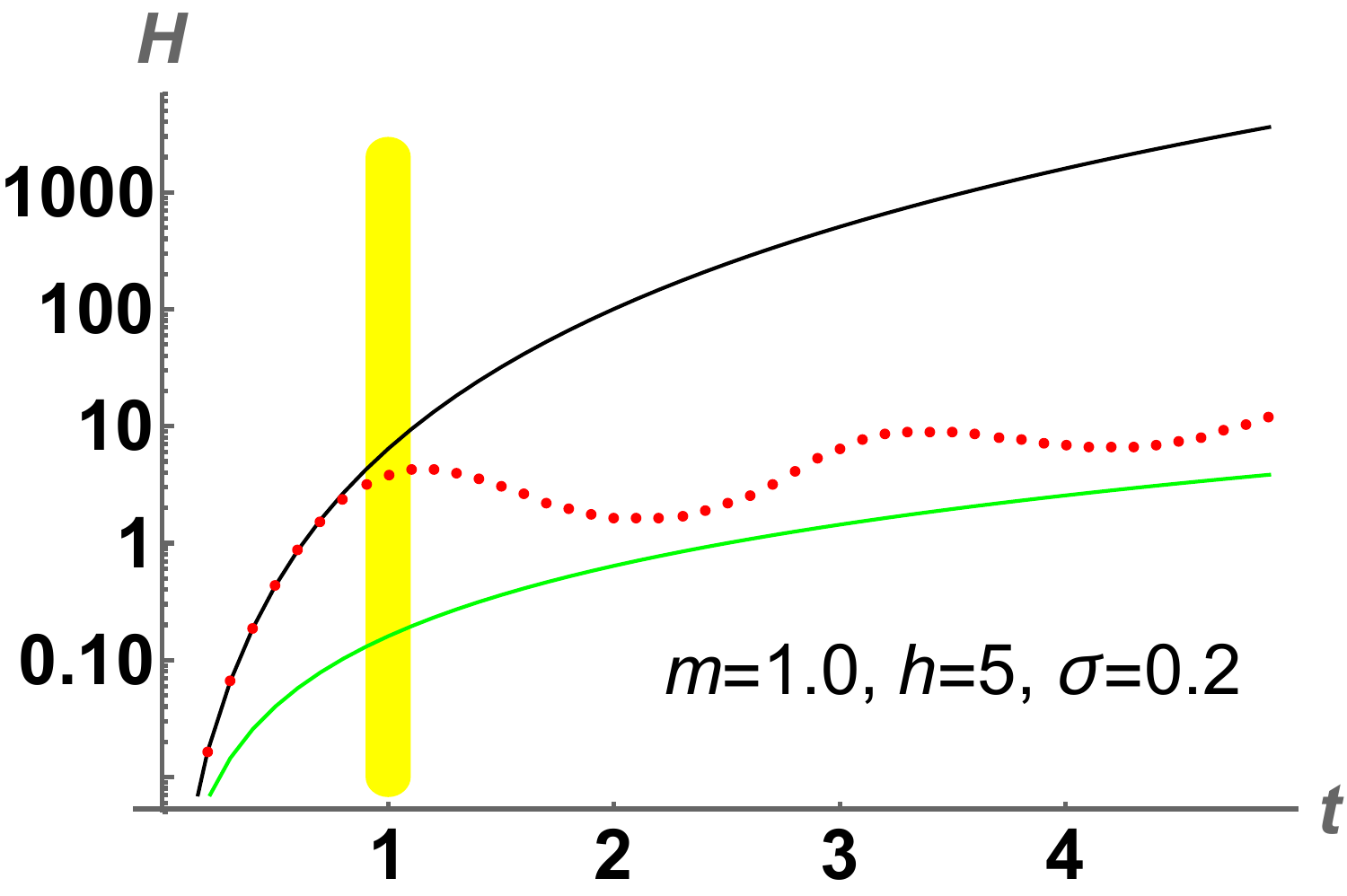}
\includegraphics[width=0.49\columnwidth]{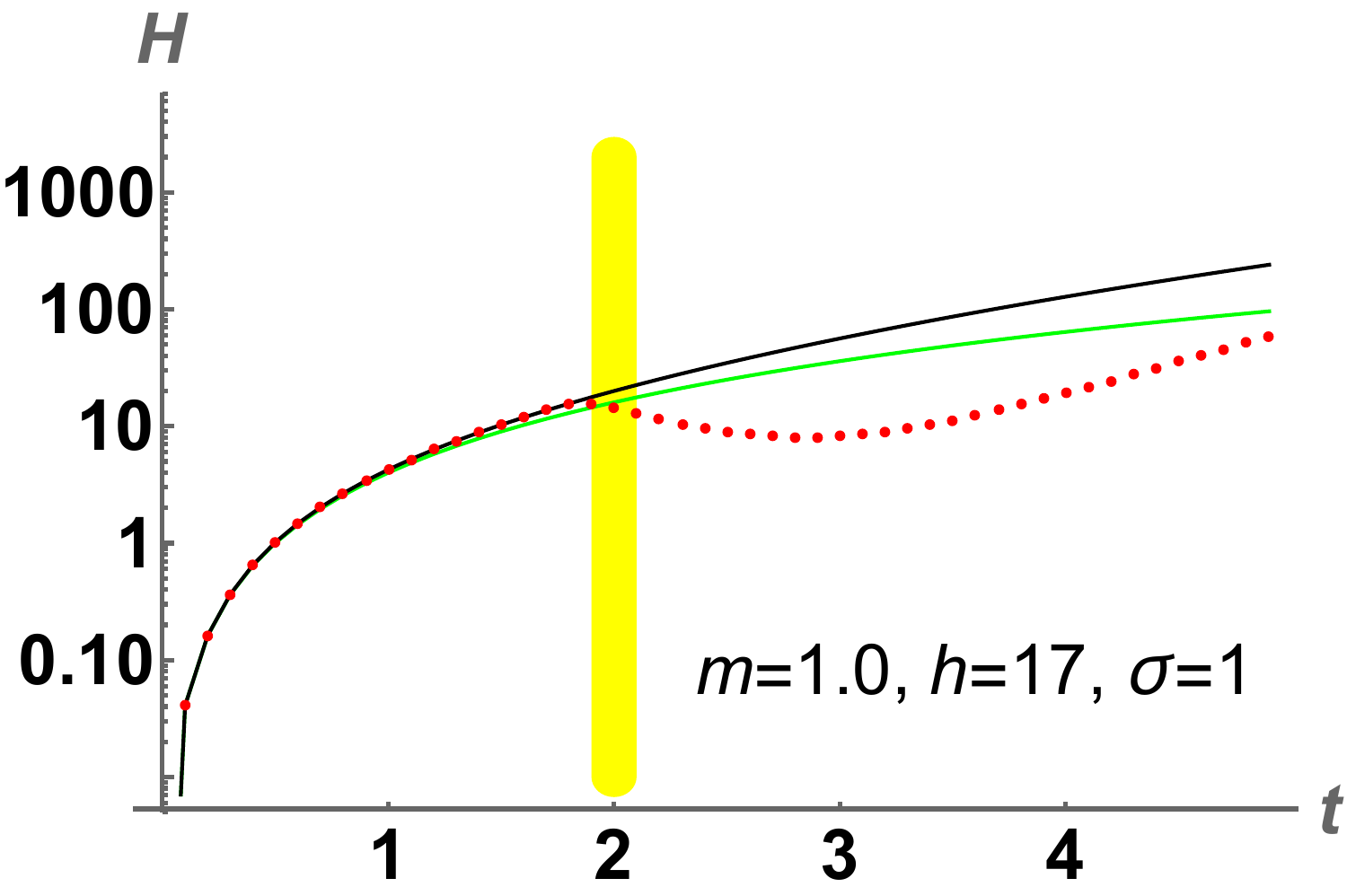}
\includegraphics[width=0.49\columnwidth]{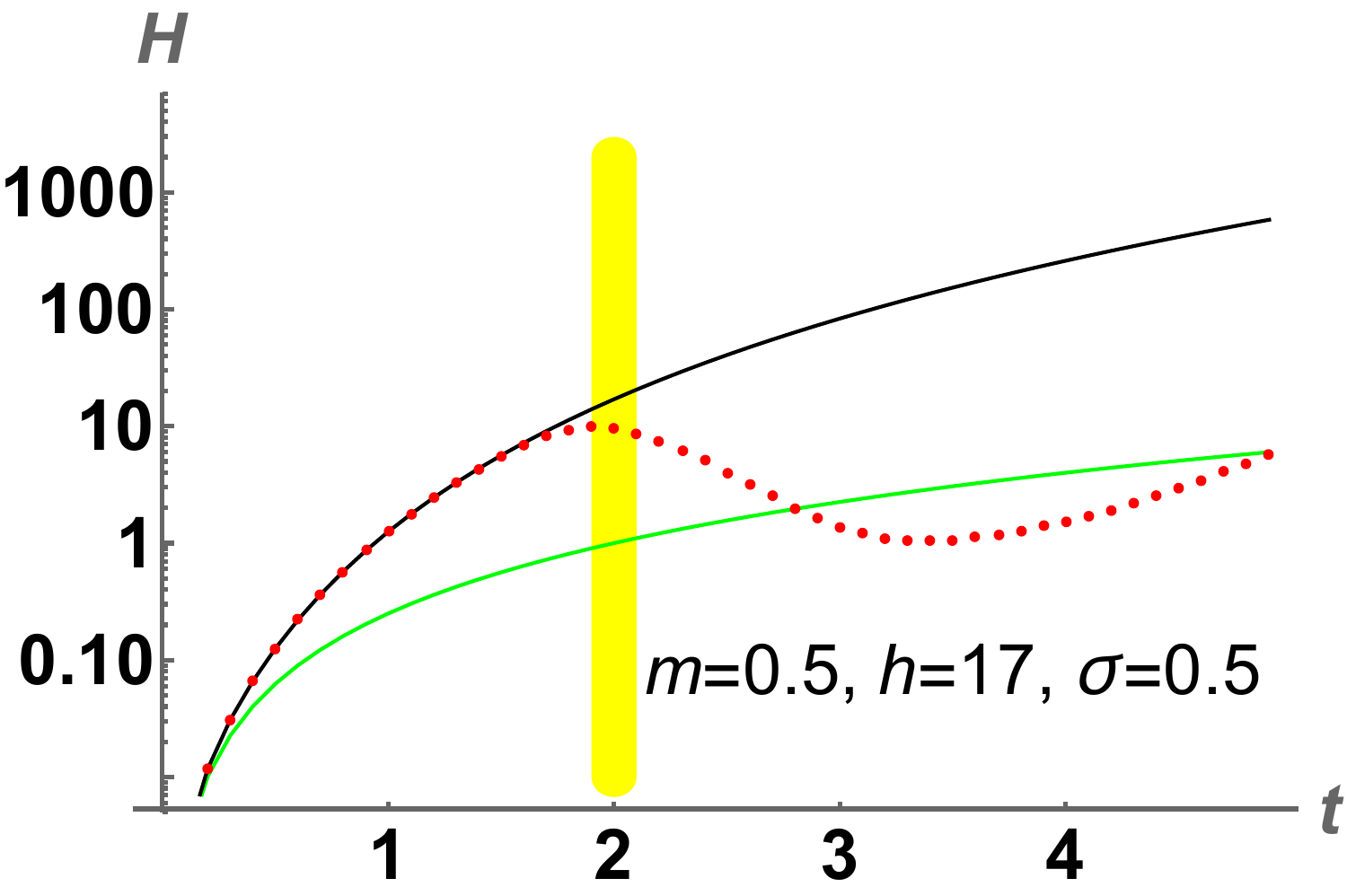}
 \caption{QFI $H(g)$ for the estimation of $g$ by a quantum bouncer (red 
  points) compared to the same quantity obtained with a free falling probe (black 
  solid lines) for different values of the involved parameters. The green line 
  denotes the $t^2$ term of the no-floor QFI. The yellow columns provide a visual 
  aid to discern the end of the range of validity of the no-floor approximation. 
\label{f:bc}}
\end{figure}

As a matter of fact, the dynamics cannot be studied analytically and we use Eqs.~(\ref{eqn:seriesexpansion}) and (\ref{eqn:coefficients}) to evaluate the 
evolved states of the quantum bouncer, and Eq.~(\ref{eqn:qfipurestate}) to calculate 
the QFI numerically. 
To contain the computational load, we focus on situations where about $\approx 100$ 
coefficients in Eq.~(\ref{eqn:singlepacket}) are enough to describe the dynamics. 
This corresponds to regimes where the combination of parameters
$k = (\frac{2m^2g}{\hbar^2})^\frac{1}{3}$ is of the order of one. The comparison 
between the QFI $H(g)$ obtained with and without considering the presence of Earth's 
surface is shown in Fig. \ref{f:bc}, for initially localized wavepackets. Similar 
results may be obtained for superpositions.  

Figure~ \ref{f:bc} displays comforting results: there is a large time range where the 
behaviour of the QFI is exactly the same as in the no-floor approximation. This means 
that until shortly before the particle hits the floor, it is as if the floor wasn't 
there at all.
\section{Estimation of $g$ by position measurement}\label{position}
We now consider a realistic measurement procedure, i.e., the measurement of the position
of the particle, that may be performed in order to estimate $g$. Our aim is to assess 
the use of superpositions and to benchmark its precision by comparing the position FI 
to the QFI. To calculate the FI, we make use of the propagator in 
Eq.~(\ref{propagatorofsysteminexam}) to find the evolved state form  
Eq.~(\ref{eqn:map}). For a localized Gaussian wavepacket we obtain
\begin{align} \label{sgwp}
&\psi(x, t) = \frac{(1+i) \sqrt{m \sigma}}{(2\pi)^{\frac14}\sqrt{t + 2 i m \sigma^2}} \\ 
\times & 
\exp\left\{-
\frac{m\left[
x^2 -x\,g t  ( t - 4i m \sigma^2 )  - \frac{1}{12}g^2 t^3 (t - 8 i m \sigma^2  )\right]}
{2(i t + 2 m \sigma^2)}\right\}
\notag
\end{align}
For position measurement, the probability distribution $p(x|g)$ 
that we need to use Eq.~(\ref{eqn:Fisher}) is immediately given by 
$|\psi(x, t)|^2$ with $\psi(x, t)$ given above. Plugging $|\psi(x, t)|^2$ and its 
derivative with respect to $g$ into Eq.~(\ref{eqn:Fisher}), we obtain the FI for 
position measurements:
\begin{equation}\label{eqn:fisherposition}
F_{x,loc}(g) = \frac{{m^2 t^4 \sigma^2}}{{t^2 + 4 m^2 \sigma^4}}\,.
\end{equation}
We notice that the dependence on the parameters of the position FI 
in Eq.~(\ref{eqn:fisherposition}) is different from that of the QFI, e.g., 
the value $\sigma = \frac{\sqrt{t}}{2\sqrt{m}}$, which minimizes the QFI, 
is now the one maximizing the FI.

Also in this case, we are interested in assessing whether using superpositions 
of localized states provides some advantages. The explicit expression of the 
position FI is clumsy and will not 
{be} reported here. Extensive numerical analysis 
shows that a convenient working regime is achieved for superpositions of highly
localized states (i.e., for small $\sigma$) while the initial momentum is not 
much relevant. 

\begin{figure}[h!]
\centering
\includegraphics[width=0.465\columnwidth]{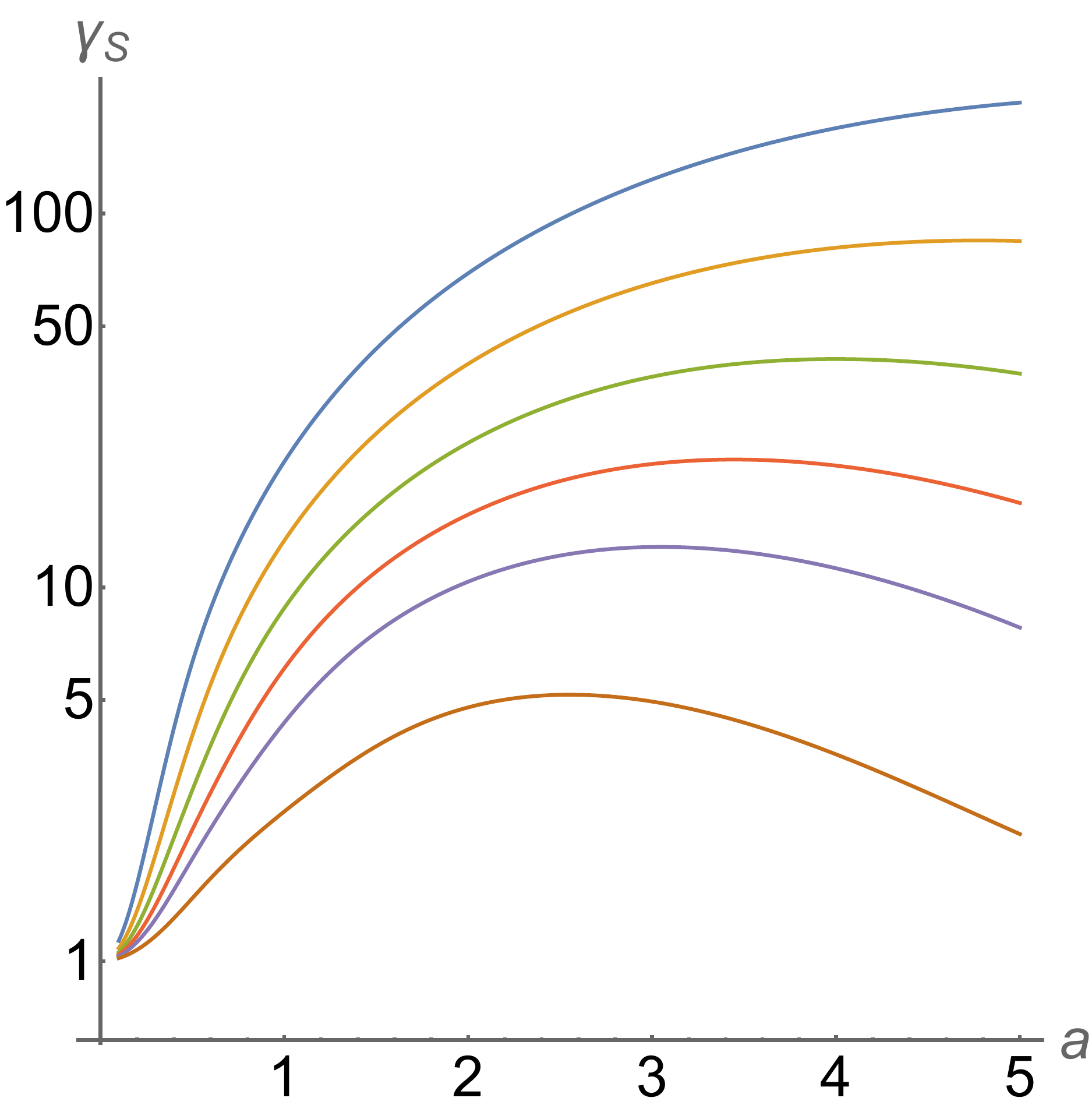} 
\includegraphics[width=0.48\columnwidth]{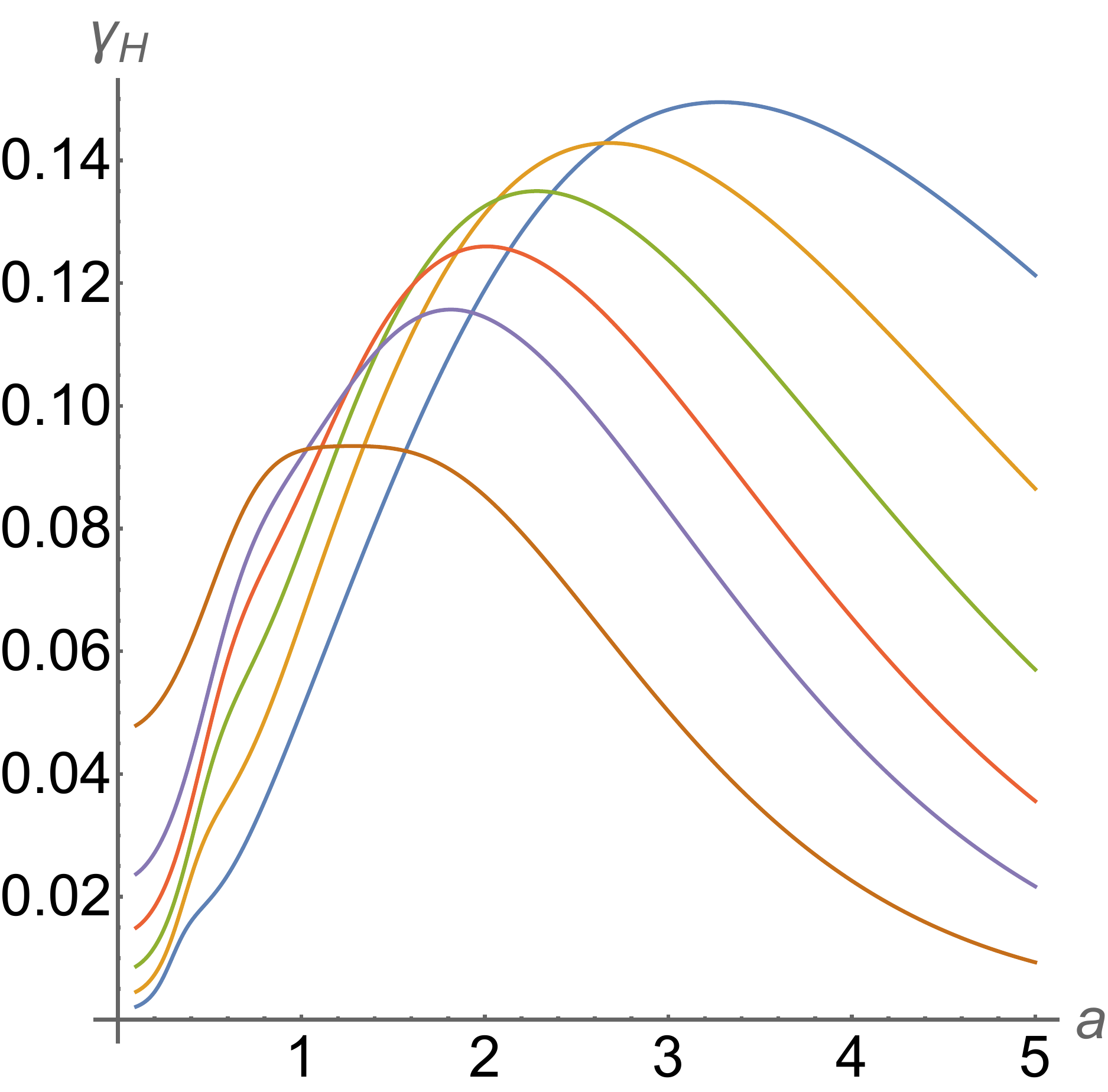}        
		\caption{(Left): the ratio $\gamma_{S}$ between the FI of position measurement 
		for probes prepared in a superposition state and that obtained for localized probes, as function of the separation $a$ and for different values of the wavepackets width $\sigma$. From top to bottom: $\sigma=0.2,0.25,0.3,0.35,0.4,0.5$.  (Right): the ratio $\gamma_{H}$ between the FI of position measurement 
		for probes prepared in a superposition state and the corresponding QFI, as function of the separation $a$ and for different values of the wavepackets width $\sigma$. From top to bottom: $\sigma=0.2,0.25,0.3,0.35,0.4,0.5$.		
		In both plots, the other parameters 
are set to $t=1.0$ and $m=0.5$ in order to safely stay within the no-floor approximation.  }
        \label{fig:fisherposition}
\end{figure}

In Fig.~\ref{fig:fisherposition} we show the two ratios 
\begin{align}
\gamma_{S} = \frac{F_{x,sup}}{F_{x,loc}} \qquad
\gamma_{H} = \frac{F_{x,sup}}{H_{sup}}
\end{align}
as a function of the separation $a$ for various values of $\sigma$. The other parameters 
are set to $t=1.0$ and $m=0.5$ in order to safely stay within the no-floor approximation. 
The ratio $\gamma_{S}$ quantifies the gain one obtains using superpositions instead of
localized probes, whereas $\gamma_{H}$ assess the precision of position measurement
against the ultimate bounds imposed by quantum mechanics. 

The plots  show that the use of superpositions is largely advantageous, 
especially for lower values of $\sigma$, and that the Fisher information of 
position measurement is a consistent fraction of the QFI. 
\section{Joint estimation of $g$ and $m$}\label{joint}
In the previous Sections, we have seen that the precision in the estimation of $g$ is 
determined by the values of the other parameters. Focusing for simplicity on the case 
of a localized probe, those parameters are the interaction time $t$, the width 
$\sigma$ of the wavepacket, and the mass of the particle $m$. The first two parameters
may be tuned by the experimenter while the mass of the particle should be determined 
independently and assumed to stay constant during the repeated preparations of the probe. 
This assumption, and the related risk of introducing systematic errors, may be avoided 
by performing the joint estimation of $g$ and $m$. In this Section, we consider the 
problem of estimating $g$ and $m$ simultaneously, and investigate whether they are 
compatible parameter \cite{he2025scrambling}, i.e., whether the joint estimation 
introduces additional 
quantum noise. 

We start by calculating the QFI matrix and the Uhlmann curvature using 
Eq.s~(\ref{eqn:Q2par}) and (\ref{eqn:Uhlmann2par}). 
\begin{align}\label{calc:Q2par}
\bm{H}(g, m) & = 4
\begin{pmatrix}
H_{loc} & \frac{g}{m}\left(H_{loc} - \frac{t^2}{4\sigma^2}\right) \\
\frac{g}{m}\left(H_{loc} - \frac{t^2}{4\sigma^2}\right) & \frac{t^2}{8 m^4 \sigma^4}+\frac{g^2 t^4}{m^2 \sigma^2}+4 g^2 \sigma^2 \\
\end{pmatrix} \\ 
\label{calc:Uhlmann2par}
\bm{D}(g, m) & = 4
\begin{pmatrix}
  0 & gt^3 \\
   -gt^3 & 0 \\
\end{pmatrix}\,.
\end{align}
Eq. (\ref{calc:Uhlmann2par}) proves that the two parameters are not compatible, since the corresponding SLDs do not commute on the relevant subspace. On the other hand, upon 
calculating the quantities $\mathcal{R}$ and $\mathcal{T}(\bm{W})$ we have 
\begin{align}\label{calc:quantumness}
\mathcal{R} & = \frac{{4\sqrt{2} g m^2 t \sigma^3}}{{\sqrt{16m^2 \sigma^4 + t^2 (1 + 32g^2 m^4 \sigma^6)}}} \\
\mathcal{T}(\bm{W}) & =  \frac{16 g m^4 t \sqrt{w} \sigma^4}{w + 2 m^2 t^2 \sigma^2 (m^2 + 4 g^2 w) + 32 m^4 \sigma^6 (m^2 + g^2 w)}\,,
\end{align}
and substituting a sensible choice of parameters (in natural units)
$g\simeq 2.15\cdot 10^{-32}$ GeV, $m\simeq 10$ a.m.u $\simeq 9.31$ GeV, $\sigma\simeq 10^{-3}\, m  \simeq 5.1\cdot 10^{-6}$ GeV$^{-1}$, we found that $\mathcal{R} \simeq \mathcal{T}(\bm{W}) < 10^{-40}$ $\forall t$ and $\forall w$. 

Additionally, by choosing $w=0$ we have $\mathcal{T}(\bm{W})=0$ and we can 
compare precision achievable in estimating $g$ knowing $m$ (given by $1/H_{loc}$) 
with that achievable without any knowledge of $m$ (treated as a nuisance parameter). 
For the same set of parameters we have $1/H_{loc} \simeq (\bm{H}^{-1})_{11}$, 
confirming that the quantum noise due to incompatibility is negligible. 
\section{Conclusions}\label{outro}
In this work, we have investigated the potential of quantum probes in enhancing the 
precision in the estimation of the gravitational acceleration constant $g$. Our results 
demonstrate that delocalized quantum probes, prepared in superposition states, 
generally outperform localized ones, with the precision enhancement scaling 
quadratically with the separation between the wavefunction components. This advantage 
persists also under realistic position measurements, where the Fisher information 
remains close to the ultimate quantum bound set by the quantum Fisher information.

We have validated the no-floor approximation for a large temporal range, showing that the 
influence of Earth's surface is negligible until shortly before the probe interacts 
with it. Furthermore, we have addressed the joint estimation of $g$ and 
of the probe mass $m$, proving that despite their inherent incompatibility as 
parameters, only negligible excess noise is introduced. This result ensures that 
simultaneous estimation does not degrade precision, making it a viable strategy 
in implementations. Our findings highlight the potential of 
quantum-enhanced sensing in gravitational measurements, 
with applications in fundamental physics and precision metrology \cite{cepollaro2023gravitational,PhysRevD.94.044019,PhysRevD.98.125007,PhysRevD.102.056012,PhysRevD.106.124035}. 

\acknowledgments
MGAP thanks Massimo Frigerio, Carlo Cepollaro, Flaminia Giacomini, Luigi Seveso, and 
Valerio Peri for interesting and useful discussions.

\appendix

\section{Quantumness of two-parameter models\label{a:twodm}}
For a generic two-parameter pure state model $| \psi_{\bm{\lambda}} \rangle $, with $\bm{\lambda} = (\lambda_1, \lambda_2)$, the QFI matrix and
the Uhlmann curvature can be evaluated via the following equations \cite{QMEP}:
\begin{align}\label{eqn:Q2par}
\bm{H(\lambda)} & = 4
\begin{pmatrix}
  a^2 + \alpha & ab + \mathrm{Re}(c) \\
   ab + \mathrm{Re}(c) & b^2 + \beta \\
\end{pmatrix}\,,
\\
\label{eqn:Uhlmann2par}
\bm{D(\lambda)} & = 4
\begin{pmatrix}
  0 & \mathrm{Im}(c) \\
   -\mathrm{Im}(c) & 0 \\
\end{pmatrix}\,,
\end{align}
where 
\begin{align}
a & = \langle \partial_{\lambda_1} \psi_{\bm{\lambda}}| \psi_{\bm{\lambda}} \rangle\,, \\
b & =  \langle \partial_{\lambda_2} \psi_{\bm{\lambda}}| \psi_{\bm{\lambda}} \rangle\,, \\
c & =  \langle \partial_{\lambda_1} \psi_{\bm{\lambda}}| \partial_{\lambda_2} \psi_{\bm{\lambda}} \rangle \,,\\
\alpha & =  \langle \partial_{\lambda_1} \psi_{\bm{\lambda}}| \partial_{\lambda_1} \psi_{\bm{\lambda}} \rangle \,, \\
\beta & =  \langle \partial_{\lambda_2} \psi_{\bm{\lambda}}| \partial_{\lambda_2} \psi_{\bm{\lambda}} \rangle\,.
\end{align}

\bibliography{gbib}   

\end{document}